\documentclass[reprint,final,floatfix,nofootinbib,superscriptaddress,longbibliography]{revtex4-2}

\usepackage{lmodern,mathtools,amssymb,mathrsfs,bbm,microtype,bm} 
\usepackage{color,graphicx} 
\definecolor{bluesmoke}{rgb}{0.207843,0.415686,0.623529}
\usepackage[
  allcolors=bluesmoke,
  colorlinks,
  pdftex,
  pdftitle={Theory of Microphase Separation in Elastomers},
  pdfauthor={Manu Mannattil, Haim Diamant, and David Andelman},
  pdfkeywords={Phase separation, pattern formation, elasticity, elastomers, phase-field modeling},
  pdfsubject={Statistical Physics, Polymers \& Soft Matter, Condensed Matter \& Materials Physics},
  hypertexnames=false
]{hyperref}
\makeatletter
\DeclareMathOperator{\sinc}{sinc}
\DeclareMathOperator{\trace}{tr}
\def\abs#1{\ensuremath{\left|#1\right|}}
\def\dd{\ensuremath{\mathrm{d}}}
\def\div{\ensuremath{\boldsymbol{\nabla\cdot}}}
\def\grad{\ensuremath{\boldsymbol{\nabla}}}
\def\kbt{\ensuremath{k_{\text{B}}}T}
\def\mr{\ensuremath{m_{\text{r}}}}
\def\ms{\ensuremath{m_{\text{s}}}}
\def\NN{\ensuremath{\nonumber}}
\def\phic{\ensuremath{\phi_{*}}}
\def\phiz{\ensuremath{\phi_{0}}}
\def\psiz{\ensuremath{\psi_{0}}}
\def\qm{\ensuremath{q_{\text{m}}}}
\def\Am{\ensuremath{A_{\text{m}}}}
\def\strain{\ensuremath{\bm{\varepsilon}}}
\def\stress{\ensuremath{\bm{\sigma}}}
\def\Tc{\ensuremath{T_{*}}}
\def\Tm{\ensuremath{T_{\text{micro}}}}
\def\trans{\ensuremath{^{\mathsf{T}}}}

\def\altsection#1{\paragraph*{#1.---\hskip-.75em}}
\def\unit{{\kern2.333pt}}
\definecolor{hlcolor}{rgb}{0.646,0.165,0.165}

\def\bibsection{%
  \par
  \begingroup
  \baselineskip26\p@
  \bib@device{\hsize}{72\p@}%
  \endgroup
  \nobreak\@nobreaktrue
  \addvspace{19\p@}%
}%

\makeatother

\begin{document}

\title{Theory of Microphase Separation in Elastomers}
\author{Manu Mannattil}
\email{manu.mannattil@posteo.net}
\affiliation{School of Chemistry, Tel Aviv University, Ramat Aviv, Tel Aviv 69978, Israel}
\affiliation{School of Physics and Astronomy, Tel Aviv University, Ramat Aviv, Tel Aviv 69978, Israel}
\affiliation{Center for Physics and Chemistry of Living Systems, Tel Aviv University, Tel Aviv 69978, Israel}
\author{Haim Diamant}
\affiliation{School of Chemistry, Tel Aviv University, Ramat Aviv, Tel Aviv 69978, Israel}
\affiliation{Center for Physics and Chemistry of Living Systems, Tel Aviv University, Tel Aviv 69978, Israel}
\author{David Andelman}
\affiliation{School of Physics and Astronomy, Tel Aviv University, Ramat Aviv, Tel Aviv 69978, Israel}
\affiliation{Center for Physics and Chemistry of Living Systems, Tel Aviv University, Tel Aviv 69978, Israel}

\begin{abstract}
Inspired by recent experiments, we present a phase-field model of microphase separation in an elastomer swollen with a solvent.
The imbalance between the molecular scale of demixing and the mesoscopic scale beyond which elasticity operates produces effective long-range interactions, forming stable finite-sized domains.
Our predictions concerning the dependence of the domain size and transition temperature on the stiffness of the elastomer are in good agreement with the experiments.
Analytical phase diagrams, aided by numerical findings, capture the richness of the microphase morphologies, paving the way to create stable, patterned elastomers for various applications.
\end{abstract}

\maketitle

\altsection{Introduction}

The interplay between elasticity and phase separation has been widely explored in various contexts since Cahn's classic work from the 1960s on spinodal decomposition~\cite{cahn1961}.
For example, a mismatch in the constituents' elastic moduli in metallic alloys can either hinder or speed up phase separation~\cite{fratzl1999}.
Similarly, elasticity regulates the morphology of the phase-separated domains in gels~\cite{rabin1997,peleg2007,dervaux2012} and liquid-crystalline fluids~\cite{loudet2000}, which can lead to intricate patterns.
Besides, mounting evidence now indicates that phase separation and elasticity are both crucial to the development of many membraneless organelles within biological cells, rekindling interest in the topic~\cite{choi2020,tanaka2022,lee2022,kumar2023,deviri2024}.
To sidestep the complexities of the biological world, several experiments have been conducted with synthetic, \emph{in vitro} model systems in the past few years~\cite{style2018,rosowski2020,rosowski2020a,wilken2023,liu2023}.
The results of these experiments, along with related theoretical work~\cite{ren2000,kothari2020,wei2020,ronceray2022,biswas2022,little2023,grabovsky2023}, once again emphasize the influence of elasticity on phase separation in soft matter systems.

A recent experiment showed elasticity-controlled microphase separation to be a highly effective technique for generating patterned elastomers with complex morphologies~\cite{fernandez-rico2024}.
In the study, a temperature quench is used to trigger microphase separation in elastomers swollen with a solvent.
The results are reminiscent of older observations of phase separation and critical density fluctuations in swollen gels as the temperature is lowered~\cite{tanaka1977,tanaka1978,li1989,onuki1993}.
In the new experiments, however, the elastomer does not fully phase separate from the solvent, and instead forms stable bicontinuous microstructures or droplets whose sizes are determined by the stiffness of the elastomer.
This microphase separation plausibly arises because of a pronounced difference in the length scales at which thermodynamics and elasticity operate~\cite{fernandez-rico2024}.
This is unlike previous examples, where patterned phases were primarily seen in systems with anisotropic elasticity or external stresses~\cite{nishimori1990} or involving nontrivial phenomena such as cavitation~\cite{kothari2020,vidal-henriquez2021}.

In this Letter, we introduce a phase-field model that captures the key features of microphase separation in swollen elastomers in the limit of weak segregation.
Recent theoretical work~\cite{qiang2024}, also inspired by the aforementioned experiments, has demonstrated that the length-scale discrepancy between elasticity and thermodynamics in elastomers can be resolved using nonlocal theories of elasticity~\cite{eringen1977,eringen1987,polizzotto2004}.
Nonlocal approaches have also been employed in other systems with scale-dependent phenomena, such as certain porous materials~\cite{derr2020} and DNA elasticity~\cite{skoruppa2021}.

The stiffness of elastomers arise primarily due to strain-induced changes in the configurational entropy of polymer chains~\cite{rubinstein2003,tanaka2011,lodge2020}.
Our scaling results for the domain size and microphase separation temperature, obtained by using results from rubber (entropic) elasticity and incorporating nonlocal effects, agree with the experimental observations.
We also highlight the diversity of the microphase morphologies by constructing a phase diagram and supplementing it with numerical results.
Put together, our findings underscore an intricate coupling between thermodynamics and elasticity, opening up novel ways to produce patternable materials for various purposes.

\altsection{Model}

\begin{figure}[b]
  \begin{center}
    \includegraphics{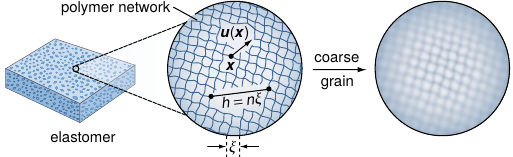}
  \end{center}
  \caption{
    Displacements $\bm{u}(\bm{x})$ occurring below a characteristic length scale $h$ do not stress the elastomer significantly.
    Using a coarse-grained strain field $\bar{\strain}$, such displacements are ``blurred away'' and filtered out.
    In our model, we choose $h$ as a multiple $n\xi$ of the end-to-end distance $\xi$ between adjacent crosslinks of the polymeric network within the elastomer.
}
  \label{main:fig:smooth}
\end{figure}

We consider a charge-neutral elastomer consisting of a crosslinked polymer network isotropically swollen with a solvent.
Polymer-solvent interaction occurs over typical intermolecular distances (e.g., the size of the solvent molecules).
On the other hand, the elastic response of the elastomer stems entirely from the underlying polymer network, which has a much larger, usually mesoscopic, characteristic length scale (Fig.~\ref{main:fig:smooth}).
Deformations of the elastomer occurring below this length scale should not engender a significant elastic response.
Elastomers can undergo large deformations during swelling, and they are customarily studied using nonlinear elasticity~\cite{dimitriyev2019}.
However, once the elastomer is completely swollen, further elastic deformations are well described using linear elasticity in terms of a three-dimensional (3D) displacement field $\bm{u}(\bm{x})$ defined over points $\bm{x}$ on the elastomer~\cite{doi2009}.
The resulting strain field is $\strain = \frac{1}{2}[\grad\bm{u} + (\grad\bm{u})\trans]$, with $(\grad\bm{u})\trans$ being the transpose of $\grad\bm{u}$.

For a precise description of thermodynamic interactions caused by compositional changes in the elastomer,
the continuum fields $\bm{u}$ and $\strain$ must both be defined at molecular length scales.
However, only those deformations occurring above a much larger length scale stress the elastomer substantially.
To address this, we consider a constitutive stress-strain relationship of the form
\begin{equation}
  \stress(\strain) = \lambda(\trace{\bar{\strain}})\mathbbm{1} + 2\mu\,\bar{\strain},
  \label{main:eq:constitutive}
\end{equation}
where $\lambda$ and $\mu$ are the Lam\'{e} parameters, $\mathbbm{1}$ is the $3\times3$ identity matrix,
and $\trace\bar{\strain}$ denotes the trace of a coarse-grained strain $\bar{\strain}$, defined by
\begin{equation}
  \bar{\strain}(\bm{x}) = \int\mathrm{d}^{3}x'\,K_{h}(\bm{x} - \bm{x}')\, \strain(\bm{x}').
  \label{main:eq:coarse}
\end{equation}
Here $K_{h}(\bm{x} - \bm{x}')$ is an isotropic, scalar kernel that depends only on the distance
$\abs{\bm{x} - \bm{x}'}$ between two points $\bm{x}, \bm{x}'$ in space.
For concreteness, we use a normalized Gaussian kernel $K_{h}({\bm{x}}) = (4\pi h^{2})^{-3/2} \mathrm{e}^{-|\bm{x}|^{2}/(4h^{2})}$, with $h$ being a suitable mesoscopic length scale that controls the extent of coarse-graining.
Nonetheless, as we demonstrate in the Supplemental Material (SM)~\cite{SM}, our results are independent of our choice for this kernel.
\nocite{gonzalez2008,han1999,douglas1993,mckenna1991,flory1976,james1944,reinitz2015,onuki1989b,jia2021,mark2004,villain-guillot1998,thiele2019,komura2000,chaikin1995,yamada2008,onuki1988a,hirotsu1991,koga1999,komura2008,hamley1997,aguirregabiria2002,bray1994,eyre1998,vollmayr-lee2003,yoon2020,github}

The stress $\stress$ computed using Eqs.~\eqref{main:eq:constitutive} and \eqref{main:eq:coarse} models the correct elastic response of the elastomer, while simultaneously allowing us to use the strain $\strain$ to capture compositional changes at molecular length scales.
This model is a particular instance of the Eringen framework~\cite{eringen1977,eringen1987,polizzotto2004} of nonlocal elasticity, and it leads to an elastic energy density $w(\strain)$ of the form
\begin{equation}
  w(\strain) = \frac{\lambda}{2}(\trace{\strain})(\trace\bar{\strain}) + \mu\trace(\strain\bar{\strain}),
  \label{main:eq:elastic_energy}
\end{equation}
obtained by contracting the strain $\strain$ with the stress $\stress$ in Eq.~\eqref{main:eq:constitutive} expressed in terms of $\bar{\strain}$.
As the kernel $K_{h}$ is positive-definite and normalized, $w(\strain)$ remains positive, bounded from below, and reduces to the usual Hookean energy density in the limit $h\to 0$.

Let the elastomer be isotropically swollen initially at a temperature $T$ with a constant volume fraction $\phiz$ of the polymer network.
Compositional changes that occur as the temperature is lowered cause the local network volume fraction $\phi(\bm{x})$ at a point $\bm{x}$ to deviate from $\phiz$.
The grand-canonical free energy of the elastomer is then given by
\begin{equation}
  \mathscr{F}[\psi, \strain] = \int\dd^{3}{x}\left[f(\psi) + \frac{1}{2}\kappa\abs{\grad\psi}^{2} + w(\strain) - \eta\psi\right].
  \label{main:eq:free}
\end{equation}
Here we have defined the order parameter (phase field) $\psi(\bm{x}) = \phi(\bm{x}) - \phic$ assuming that the homogeneous system has a ``critical'' point $(\phic,\Tc)$, and take the free-energy density $f(\psi)$ to be in a Landau form
\begin{equation}
  f(\psi) = \frac{1}{2}a(T-\Tc)\psi^{2} + \frac{1}{4}b\psi^{4},
  \label{main:eq:landau}
\end{equation}
with $a$ and $b$ being positive phenomenological constants.
Contributions from polymer-solvent mixing are included in this phenomenological $f(\psi)$, with the $\psi^{4}$ term playing an additional role in stabilizing phase separation.
For polymer networks crosslinked in solution, the critical temperature $\Tc$ is likely to be close to the theta temperature before crosslinking~\cite{horkay2023}.
Similar free-energy densities have been used to model swelling and deswelling of gels~\cite{onuki1999,onuki2002,yamaue2000,yamaue2004}.
Also included in Eq.~\eqref{main:eq:free} is the elastic energy density $w(\strain)$ and a gradient-squared term
with an interfacial parameter $\kappa >0$ to penalize spatial variations in $\psi$.
Finally, $\eta$ is a Lagrange multiplier
to constrain the mean value of $\psi$ to $\psiz = \phiz - \phic$, thereby conserving the total volume of the polymer network.

For small deformations of the elastomer close to the critical point, the strain $\strain$ and the order parameter $\psi$ are related by a material conservation relation (SM~\cite{SM}),
\begin{equation}
  \trace\strain = \div\bm{u} \approx -\phic^{-1}\psi.
  \label{main:eq:matcon}
\end{equation}
Compositional changes in the polymer volume fraction during temperature quenches arise primarily via solvent diffusion.
This allows us to disregard shear deformations and use Eqs.~\eqref{main:eq:elastic_energy} and~\eqref{main:eq:matcon} to write the total elastic energy as a binary interaction in $\psi$ mediated by the coarse-graining kernel $K_{h}$.

For linear stability analysis of Eq.~\eqref{main:eq:free}, we express the order parameter ${\psi}(\bm{x})$ and the kernel $K_{h}(\bm{x})$ in terms of their Fourier transforms,
$\psi_{\bm{q}} = \int \mathrm{d}^{3}x\, \mathrm{e}^{-i\bm{q}\cdot\bm{x}}\, \psi(\bm{x})$ and $K_{h}(\bm{q}) = \mathrm{e}^{-h^{2}q^{2}}$.
Upon expressing the quadratic part of the total free energy in Fourier space, we find (SM~\cite{SM})
\begin{equation}
  \mathscr{F}[\psi] = \frac{1}{2}\int\frac{\mathrm{d}^{3}q}{(2\pi)^{3}}\psi_{-\bm{q}}F_{\bm{q}}\psi_{\bm{q}} + \int\mathrm{d}^{3}x\left(\frac{1}{4}b\psi^{4} - \eta\psi\right),
  \label{main:eq:free_fourier}
\end{equation}
where $F_{\bm{q}}$ is the Fourier transform of the effective binary interaction for $\psi$ given by
\begin{equation}
  F_{\bm{q}} = a(T-\Tc) + \kappa q^{2} + M\mathrm{e}^{-h^{2}q^{2}}.
  \label{main:eq:interaction}
\end{equation}
Here $q = \abs{\bm{q}}$ and $M = (\lambda + 2\mu)/\phic^{2}$ is the rescaled longitudinal modulus~\cite{rabin1994,doyle2021} of the swollen elastomer.

The second term in Eq.~\eqref{main:eq:interaction}, which favors long-range (small $q$) modulations in $\psi$, measures the energy cost to create interfaces.
Meanwhile, the elastic term $M\mathrm{e}^{-h^{2}q^{2}}$ favors short-range (large $q$) modulations.
Hence, we expect the emergence of a stable, spatially modulated phase at an intermediate length scale, provided that the elastic term is adequately large.
The characteristic size of the modulated phase scales as $\Lambda \sim 2\pi\qm^{-1}$, where $\qm$ is the wavenumber at which $F_{\bm{q}}$ acquires its minimum.

From Eq.~\eqref{main:eq:interaction}, we see that $F_{\bm{q}}$ has a minimum at a nonzero $\qm$ given by
\begin{equation}
  q_{\mathrm{m}}^{2} = h^{-2}\ln\gamma,
  \label{main:eq:qm}
\end{equation}
only if the dimensionless parameter $\gamma = Mh^{2}/\kappa > 1$.
The parameter $\gamma$, which measures the relative importance of elastic and interfacial effects,
is analogous to the (inverse) elastocapillary number~\cite{liu2019,ronceray2022}
and the Lifshitz point~\cite{hornreich1975} of Eq.~\eqref{main:eq:free_fourier} is located at $\gamma = 1$ (SM~\cite{SM}).
If the elastic energy cost exceeds the cost to form interfaces ($\gamma > 1$), the system can minimize its total energy by creating many stable, finite-sized domains, resulting in microphase separation.
Note that if $h = 0$ in Eq.~\eqref{main:eq:coarse} and the system exhibits a local elastic response, we can recover known results for swollen polymer networks from the free energy in Eq.~\eqref{main:eq:free_fourier}, such as the onset of spinodal decomposition at temperatures where the osmotic longitudinal modulus vanishes, negative Poisson's ratio etc.~\cite{onuki1988} (SM~\cite{SM}).

During a temperature quench from the uniform phase with $\psi(\bm{x}) = \psiz$, the onset of microphase separation is indicated by linear instability in the order-parameter fluctuations.
Upon expressing $\psi(\bm{x}) = \psiz + \delta\psi(\bm{x})$ and expanding the free energy in Eq.~\eqref{main:eq:free_fourier} up to $\mathcal{O}(\delta\psi^{2})$ in the fluctuations $\delta\psi(\bm{x})$, we determine that the instability arises at a temperature where $F_{\bm{q}} = -3b\psiz^{2}$ and $q = \qm$.
This provides an estimate for the temperature $\Tm$ at which microphase separation begins, which we find to be
\begin{equation}
  \Tm(\psiz) = \Tc - a^{-1}\left[3b\psiz^{2} + M\gamma^{-1}\left(1+\ln\gamma\right)\right].
  \label{main:eq:tm}
\end{equation}
Clearly, $\Tm$ decreases linearly with the modulus $M$, showing that deeper temperature quenches are required to induce microphase separation in stiffer elastomers.

\altsection{Comparison to experiments}

Polydimethylsiloxane (PDMS) elastomers, such as the ones used in the experiments in Ref.~\cite{fernandez-rico2024}, are susceptible to chain entanglement effects that can alter their elastic response substantially, particularly at low crosslink densities~\cite{gaylord1987,gaylord1990,douglas2010,douglas2024}.
However, based on the observed variation of the Young's modulus with crosslink density (detailed in the SM~\cite{SM}), we judge entanglement effects to be negligible, enabling us to use classical rubber elasticity theory in our analysis.

Apart from the intermolecular distance, a relevant length scale in elastomers is the root-mean-square end-to-end distance $\xi$ of the strands between adjacent crosslinks in the polymer network~\cite{canal1989,yoo2006,parrish2017,richbourg2020} (Fig.~\ref{main:fig:smooth}).
Taking each strand to be a freely-jointed chain with a Flory ratio $C_{\infty}$~\cite{rubinstein2003}, composed of $N$ repeat units of length $\ell$, we have
$\xi^{2} \sim \frac{1}{2}C_{\infty}N\ell^{2}$~\cite{tanaka2011,richbourg2020}.
Here the factor of $\frac{1}{2}$ is an estimate assuming the network junctions have tetrafunctional connectivity.
If the strands and the repeat units have molecular masses $\ms$ and $\mr$, respectively, then $N = \ms/\mr$.
Assuming that the elastomer has a mass density $\rho$, its Young's modulus in the dry state is $Y = 3\rho\kbt/2\ms$~\cite{lodge2020}.
Using this expression to write $\ms$ and $N$ in terms of $Y$, we find that the end-to-end distance scales as~\cite{yoo2006,parrish2017}
\begin{equation}
  \xi \sim (3B/Y)^{1/2},
  \label{main:eq:xi}
\end{equation}
where $B = C_{\infty}\rho\ell^{2}\kbt/(4\mr)$ is a material-dependent parameter, with $k_{\text{B}}$ being the Boltzmann constant.
See the SM~\cite{SM} for further details.
For PDMS elastomers we find $B = 0.006\unit\text{kPa}\unit\text{\textmu m}^{2}$, which gives $\xi \sim 5\text{--}50\unit\text{nm}$ for the experimental range of $Y \sim 10\text{--}800\unit\text{kPa}$.
Compared to $\xi$, the intermolecular length scale ($\sim \ell$) is of the order of a few \r{A}.
The polymer network within the elastomer can be treated as an elastic continuum only at length scales much larger than $\xi$, but proportional to it.
For this reason, we take the coarse-graining length scale to be $h = n\xi$.
Here, the phenomenological factor $n$ can be interpreted as the average number of crosslinks we coarse-grain over in each direction (Fig.~\ref{main:fig:smooth}).
Its value depends on the kernel used in Eq.~\eqref{main:eq:coarse}, with wider, long-range kernels giving smaller values for $n$.

\begin{figure}
  \begin{center}
    \includegraphics{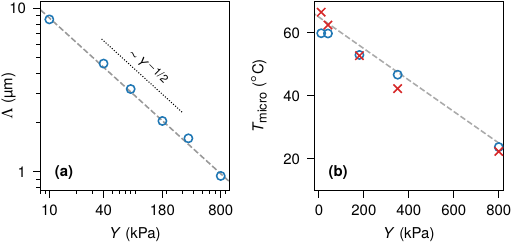}
  \end{center}
  \caption{(a)~Domain size $\Lambda$ as a function of the Young's modulus $Y$ of the dry elastomer (log-log plot).
    The circles indicate experimental values of $\Lambda$ for PDMS elastomers from Ref.~\cite{fernandez-rico2024}, showing the scaling $\Lambda \sim Y^{-1/2}$.
    The dashed line represents the prediction from Eq.~\eqref{main:eq:domain_size} with $\kappa = 0.013\unit\text{kPa}\unit\text{\textmu m}^{2}$ and fitting parameters $n = 110$, $\phic=0.2$.
    (b)~Decrease in the microphase separation temperature $\Tm$ with $Y$.
    The circles show the experimental values of $\Tm$ for an initial swelling temperature of $60\unit^{\circ}\text{C}$~\cite{fernandez-rico2024}.
    The crosses represent $\Tm$ estimated from Eq.~\eqref{main:eq:tm} using experimental values of the mean polymer volume fraction $\phiz$, with the dashed guideline illustrating the linearity between $\Tm$ and $Y$.
    Other fitting parameters are $a=0.025\unit\text{kPa}\unit\text{K}^{-1}$, $b = 2\unit\text{kPa}$, and $\Tc = 70\unit^{\circ}\text{C}$.
  }
\label{main:fig:comparison}
\end{figure}

In order to estimate the domain size $\Lambda$ of the microphases, we note that the rescaled longitudinal modulus $M$ of a swollen elastomer is related to its dry Young's modulus $Y$ via $M \sim \frac{1}{3}\phic^{-5/3}Y$~\cite{onuki1993,SM}.
Interface formation occurs at intermolecular length scales, so we estimate the interfacial parameter as $\kappa \sim \kbt/\ell$~\cite{leibler1987}.
With the choice $h = n\xi$, the parameter $\gamma$ is independent of $Y$, and using Eqs.~\eqref{main:eq:qm} and~\eqref{main:eq:xi} we find the scaling
\begin{equation}
  \Lambda \sim 2\pi\left[\frac{3Bn^{2}}{Y\ln\left(Bn^{2}\kappa^{-1}\phic^{-5/3}\right)}\right]^{1/2}\hspace{-1em}.
  \label{main:eq:domain_size}
\end{equation}
The scaling $\Lambda \sim Y^{-1/2}$ above is markedly different from what one would expect on dimensional grounds alone (the so-called rheological mesh size of polymer networks that scales as $(\kbt/Y)^{1/3}$~\cite{wisniewska2018,tsuji2018}).
In Fig.~\ref{main:fig:comparison}(a), we compare the experimental results and Eq.~\eqref{main:eq:domain_size} and find good agreement between the two.
Furthermore, as we see from Fig.~\ref{main:fig:comparison}(b), the microphase separation temperature $\Tm$ linearly decreases with $Y$, which is consistent with the prediction of Eq.~\eqref{main:eq:tm}.
Using the $\Tm$ data, one can estimate the parameters $(\phic, \Tc)$ appearing in Eq.~\eqref{main:eq:landau}.
We show in the SM~\cite{SM} that these scalings for $\Lambda$ and $\Tm$ are agnostic to the choice of the kernel $K_{h}$ in Eq.~\eqref{main:eq:coarse}.

\begin{figure}
  \begin{center}
    \includegraphics{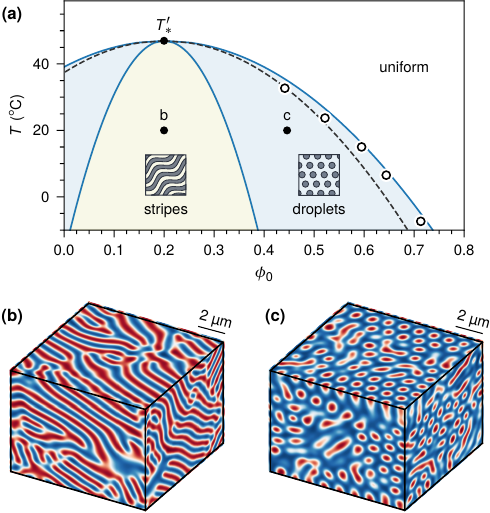}
  \end{center}
  \vspace{-0.75em}
  \caption{
    (a) Phase diagram in the $(\phiz,T)$ plane for an elastomer with a dry Young's modulus $Y = 800\unit\text{kPa}$.
    Here $T$ is the temperature, and $\phiz$ is the mean polymer volume fraction.
    Other parameters are the same as in Fig.~\ref{main:fig:comparison}.
    The solid curves show the phase boundaries (binodals).
    Phase coexistence regions are not depicted as they are very narrow.
    The dashed curve depicts the microphase separation temperature $\Tm(\psi)$ from Eq.~\eqref{main:eq:tm} with a shifted critical temperature $\Tc' = \Tm(0)$.
    The open circles represent experimental results from Ref.~\cite{fernandez-rico2024}.
    (b),~(c)~Equilibrium morphologies of the elastomer obtained by numerically minimizing the free energy, Eq.~\eqref{main:eq:free_fourier}, with the corresponding $(\phiz, T)$ values marked in (a).
    Solvent-rich ($\phi < \phiz$) and solvent-deficient ($\phi > \phiz$) regions are highlighted in red and blue, respectively.
    \vspace{-0.75em}
  }
  \label{main:fig:phase}
\end{figure}

Close to the critical point, the microphase domain boundaries are diffuse (weak segregation), and they are well approximated as modulations in the order parameter $\psi$ with a wavenumber $q = \qm$~\cite{seul1995,andelman2009}.
A phase diagram in the $(\phiz, T)$ plane constructed using this one-mode approximation is presented in Fig.~\ref{main:fig:phase}(a), with the analytical steps detailed in the SM~\cite{SM}.
For simplicity, we have only examined 2D modulations
in the phase diagram.
Nonetheless, it shows excellent agreement with the equilibrium phases found by numerically minimizing the free energy in 3D [Figs.~\ref{main:fig:phase}(b) and \ref{main:fig:phase}(c)].
Near the critical point, there are three distinct phases: a uniform phase,
a droplet (hexagonal) phase consisting of solvent-rich droplets embedded within the elastomer,
and a stripe phase composed of alternating solvent-rich and solvent-deficient layers.
An ``inverted'' droplet phase also appears at low $\phiz$.

In Fig.~\ref{main:fig:phase}(a), the first-order phase-transition curves that divide the different phases converge at a critical point $\Tc'$, where a second-order transition between the uniform and the stripe phase is possible.
The phase diagram has the same topology as phase diagrams for block copolymers~\cite{leibler1980,fredrickson1987} and other systems displaying modulated phases~\cite{garel1982,andelman1987,elder2004,provatas2011}, which are often characterized by a Landau--Brazovskii free energy.
We show in the SM~\cite{SM} that the free energy in Eq.~\eqref{main:eq:free_fourier} can be simplified to this form, explaining the generic nature of the phase diagram, which also has regions of phase coexistence~\cite{SM}.
However, for the experimental parameter ranges used here, the widths of these regions are very small, and therefore are not depicted.
The absence of substantial regions of phase coexistence may account for the apparent lack of hysteresis seen in the experiments~\cite{fernandez-rico2024}.

The phase diagram in Fig.~\ref{main:fig:phase}(a) shows good agreement with the experimental results and predicts the onset of microphase separation well.
Experimentally, droplets are seen in soft elastomers with $Y \lesssim 40\unit\text{kPa}$.
Only bicontinuous structures (different from stripes and droplets) are observed in stiffer elastomers.
However, because of the generic topology of the theoretical phase diagram, irrespective of the stiffness, we expect the droplet phase to always appear first during an off-critical temperature quench.
This suggests that some other mechanism is responsible for the emergence of bicontinuous structures in stiffer samples, e.g., shear deformations or nonlinear effects, which we have neglected.
Further consistency with experiments is seen upon examining the static structure factor, found using Eq.~\eqref{main:eq:free_fourier} as $S(q) \sim F_{\bm{q}}^{-1}$.
It peaks at $q = \qm$, given in Eq.~\eqref{main:eq:qm}, and explains the smooth increase of the scattering intensity at a fixed $q$ during a temperature quench as seen in the experiments (SM~\cite{SM}).

\altsection{Summary and outlook}

Using a phase-field model for swollen elastomers, we have predicted the possibility of a microphase separation
arising from an imbalance between the intermolecular length scale and the mesoscopic coarseness of network elasticity.
The elastomer remains stable with an intrinsically selected length scale if the free-energy contribution from the network elasticity is adequately large compared to the interfacial energy costs.
Our scaling predictions for the domain size $\Lambda$ and the microphase separation temperature $\Tm$ as a function of the elastic moduli are consistent with recent experimental observations~\cite{fernandez-rico2024}.

As the number of repeat units $N$ between the crosslinks follows the scaling $N \sim Y^{-1}$, we find $\Lambda \sim N^{1/2}$ and a linear dependence between $\Tm$ and $N^{-1}$.
Intriguingly, similar scaling behaviors have been experimentally observed in crosslinked polymer blends~\cite{briber1988,read1995} and were predicted earlier by \citet{gennes1979} using a phenomenological model that draws an analogy to electrostatics.
The similarity in the scaling suggests that the internal elastic response of these blends may be nonlocal.

Our use of nonlocal elasticity was motivated by a recent theoretical study~\cite{qiang2024} inspired by the same experiments on elastomers.
In this study, a one-dimensional (1D) nonlocal model was used to obtain the scaling $\Lambda \sim Y^{-1/2}h^{1/2}\kappa^{1/4}$ in the strong-segregation limit, taking the nonlocality scale $h$ and the Young's modulus $Y$ to be independent.
This scaling is different from our 3D result for weak segregation, Eq.~\eqref{main:eq:domain_size}, which also takes into account the inter-dependence of $h$ and $Y$, incorporating results from rubber elasticity.
Furthermore, our model predicts a first-order transition from the uniform phase to various patterned phases.
Conversely, in the 1D nonlocal model, a line of second-order transitions was predicted for large $Y$ based on detailed numerical analyses~\cite{qiang2024}.
Differences in the dimensionalities of the two models may explain this discrepancy (SM~\cite{SM}).

Microphase separation in elastomers closely resembles that in block copolymers and other systems exhibiting modulated phases~\cite{seul1995}.
  As we discuss in the SM~\cite{SM}, we expect it to be a rich source of related phenomena such as fluctuation-induced first-order transitions~\cite{brazovskii1975,fredrickson1987}, Lifshitz behavior~\cite{bates1995,bates1997}, microemulsion phases, etc.~\cite{teubner1987,komura2007}.
  Other experimentally relevant theoretical questions include the kinetics of phase separation~\cite{onuki1999,li1995}, effect of quenched impurities and network heterogeneities~\cite{zhao2009,ghosh2020,ghosh2024}, volume phase transitions~\cite{dusek2020,mussel2021}, etc. Finally, extensions of our theory to ternary systems in the strong segregation regime could help elucidate the non-power-law scaling of the domain size observed in earlier experiments~\cite{style2018}.

\altsection{Acknowledgments}

We thank Ram Adar, Enrico Carlon, Amit Kumar, Chengjie Luo, L. Mahadevan, Yicheng Qiang, Yitzhak Rabin, Sam Safran, Lev Truskinovsky, and David Zwicker for useful conversations.
M.M.~acknowledges partial support through the Bloomfield Fellows Program at Tel Aviv University.
H.D.~acknowledges support from the Israel Science Foundation (ISF Grant No.~1611/24).
D.A.~acknowledges support from the Israel Science Foundation (ISF Grant No.~226/24).

\bibliography{library,misc}

\makeatletter
\clearpage
\pagebreak
\widetext
\def\set@footnotewidth{\onecolumngrid}
\def\footnoterule{\hrule width 75pt\relax}
\setlength{\skip\footins}{\baselineskip}

\setcounter{equation}{0}
\setcounter{figure}{0}
\setcounter{footnote}{0}
\setcounter{mpfootnote}{0}
\setcounter{page}{1}
\setcounter{paragraph}{0}
\setcounter{part}{0}
\setcounter{section}{0}
\setcounter{subparagraph}{0}
\setcounter{subsection}{0}
\setcounter{subsubsection}{0}
\setcounter{table}{0}

\def\theequation{S\arabic{equation}}
\def\thefigure{S\arabic{figure}}
\def\thepage{S\arabic{page}}
\makeatother

\begin{center}
  \large \textbf{Supplemental Material:\\
  Theory of Microphase Separation in Elastomers}\\[1.25em]
  \normalsize Manu Mannattil,$^{1,2,3}$ Haim Diamant,$^{1,3}$ and David Andelman$^{2,3}$\\[0.25em]
  $^{1}$\small\emph{School of Chemistry, Tel Aviv University, Ramat Aviv, Tel Aviv 69978, Israel}\\
  $^{2}$\small\emph{School of Physics and Astronomy, Tel Aviv University, Ramat Aviv, Tel Aviv 69978, Israel}\\
  $^{3}$\small\emph{Center for Physics and Chemistry of Living Systems, Tel Aviv University, Tel Aviv 69978, Israel}
\end{center}
\vspace{1.5em}

This Supplemental Material is organized as follows:
In Sec.~\ref{sec:swollen}, we derive an expression for the free energy of swollen elastomers.
A comparison of recent experimental results with our model is presented in Sec.~\ref{sec:comparison}.
Phase diagrams are discussed in Sec.~\ref{sec:phase}, where we also compare our model to other pattern-forming systems.
The effects of fluctuations, scattering properties, and Lifshitz behavior are presented in Sec.~\ref{sec:fluctuations}.
In Sec.~\ref{sec:coarse}, we describe a general coarse-graining procedure and prove that the domain size scaling is independent of the specific choice of the coarse-graining kernel.
Finally, in Sec.~\ref{sec:numerical}, we discuss the numerical techniques used in this work.

\section{Free energy of swollen elastomers}
\label{sec:swollen}

\subsection{Elastic deformation and material conservation}

Consider an isotropically swollen elastomer with a constant polymer network volume fraction $\phiz$ occupying a finite volume in 3D and described by the Lagrangian (material) coordinates $\bm{x}$.
As the elastomer gets deformed, the coordinate $\bm{x}$ gets transformed to the Eulerian (spatial) coordinate $\bm{y}$, and the local polymer volume fraction changes from $\phiz$ to $\phi(\bm{y})$.
We assume that the deformation is captured by a smooth invertible map $\chi$ as
\begin{equation}
  \bm{y} = \chi(\bm{x}) = \bm{x} + \bm{u}(\bm{x}),
\end{equation}
where we have introduced the displacement field $\bm{u}(\bm{x})$ (see Fig.~\ref{main:fig:smooth} of the Letter).
Material conservation within an arbitrary subvolume $\Omega$ before and after deformation gives~\cite{gonzalez2008}
\begin{equation}
  \int_{\Omega} \mathrm{d}^{3}x\, \phiz =
  \int_{\chi(\Omega)} \mathrm{d}^{3}y\, \phi(\bm{y}) =
  \int_{\Omega} \mathrm{d}^{3}x\,\abs{\det\grad\chi}\, \phi(\bm{x}).
  \label{eq:matint}
\end{equation}
In the last step of Eq.~\eqref{eq:matint}, we have changed variables back to $\bm{x}$ and write $\phi(\chi(\bm{x}))$ as just $\phi(\bm{x})$ for simplicity.
This introduces the Jacobian factor $\det(\grad \chi)$, which is always positive as physical deformations preserve orientation.
Because of the arbitrariness of $\Omega$, we equate the first and last integrands in Eq.~\eqref{eq:matint} to obtain the local material conservation relation
\begin{align}
  \phiz = \phi(\bm{x})\det(\grad\chi) &= \phi(\bm{x})\det\left(\mathbbm{1} + \grad\bm{u}\right)\NN\\
                                      &= \phi(\bm{x})\left[1 + \div\bm{u} + \mathcal{O}(\abs{\bm{u}}^{2})\right].
\end{align}
For small deformations close to the critical point, we can expand both $\phiz$ and $\phi(\bm{x})$ around $\phi = \phic$ and express $\div\bm{u}$ as
\begin{equation}
  \div\bm{u} = -\phic^{-1}\psi + \mathcal{O}(\psiz) + \mathcal{O}(\psi^{2}).
  \label{eq:matcon}
\end{equation}
Here $\psiz = \phiz - \phic$ and $\psi(\bm{x}) = \phi(\bm{x}) - \phic$.
This completes the derivation of Eq.~\eqref{main:eq:matcon} of the Letter.

\subsection{Energy of a swollen elastomer under small deformations}

The Fourier transforms of the strain $\strain$ and the coarse-grained strain $\bar{\strain}$ are
\begin{equation}
  \varepsilon_{jk}(\bm{q}) = \frac{i}{2}\left[q_{j}u_{k}(\bm{q}) + q_{k}u_{j}(\bm{q})\right]
  \quad\text{and}\quad
  \bar{\varepsilon}_{jk}(\bm{q}) = \frac{i}{2}\left[q_{j}u_{k}(\bm{q}) + q_{k}u_{j}(\bm{q})\right]K_{h}(\bm{q}),
\end{equation}
where $\bm{u}_{\bm{q}} = \int\dd^{3}x\,\mathrm{e}^{-i\bm{q}\cdot\bm{x}} \bm{u}(\bm{x})$ and $K_{h}(\bm{q})$ are the Fourier transforms of the displacement field $\bm{u}(\bm{x})$ and the coarse-graining kernel $K_{h}(\bm{x})$, respectively.%
\footnote{For subscripted terms such as $K_{h}(\bm{x)}$, we write the Fourier transform as $K_{h}(\bm{q})$ to avoid ambiguity.}
Making use of the Parseval--Plancherel identity, we find the elastic energy $\mathscr{F}_{\text{el}}$ in Fourier space to be
\begin{equation}
  \mathscr{F}_{\text{el}}[\bm{u}_{\bm{q}}] = \frac{1}{2}\int \frac{\dd^{3}q}{(2\pi)^{3}}K_{h}(\bm{q})\Big[(2\mu
  + \lambda)(\bm{q}\cdot\bm{u}_{\bm{q}})(\bm{q}\cdot\bm{u}_{-\bm{q}}) + \mu q^{2}\bm{u}^{\perp}_{\bm{q}}\cdot\bm{u}^{\perp}_{-\bm{q}}\Big],
\end{equation}
where $\bm{u}_{\bm{q}}^{\perp} = \bm{u}_{\bm{q}} - (\hat{\bm{q}}\cdot\bm{u}_{\bm{q}})\hat{\bm{q}}$ is the transverse component of $\bm{u}_{\bm{q}}$ with $\hat{\bm{q}} = q^{-1}\bm{q}$.
After Fourier transforming the material conservation relation, Eq.~\eqref{eq:matcon}, which relates $\bm{u}$ and $\psi$,
we can write the elastic energy (up to additive constants) in Fourier space as
\begin{equation}
  \mathscr{F}_{\text{el}}\left[\psi_{\bm{q}}, \bm{u}_{\bm{q}}^{\perp}\right] = \frac{1}{2}M\int \frac{\dd^{3}q}{(2\pi)^{3}}\,K_{h}(\bm{q})\psi_{\bm{q}}\psi_{-\bm{q}} +
\frac{1}{2}\mu\int\frac{\dd^{3}q}{(2\pi)^{3}}\,q^{2}K_{h}(\bm{q})\bm{u}^{\perp}_{\bm{q}}\cdot\bm{u}^{\perp}_{-\bm{q}}.
\end{equation}
Here $M = \phic^{-2}(\lambda + 2\mu)$ is the elastic part of the rescaled longitudinal (or the pressure-wave) modulus~\cite{rabin1994}.
During phase separation, compositional changes in the polymer volume fraction arise primarily via solvent diffusion,
which is not expected to affect the transverse shear modes $\bm{u}^{\perp}$.
This allows us to discard them and write the total free energy as
\begin{align}
  \mathscr{F}[\psi] &= \mathscr{F}_{\text{el}} + \int\dd^{3}x\,\left[f(\psi) + \frac{1}{2}\kappa\abs{\grad\psi}^{2}\right]\NN\\
                    &= \frac{1}{2}\int\frac{\dd^{3}q}{(2\pi)^{3}}\left[a(T-\Tc) + \kappa q^{2} + MK_{h}(\bm{q})\right]\psi_{\bm{q}}\psi_{-\bm{q}} + \int \dd^{3}x\,\left(\frac{1}{4}b\psi^{2}-\eta\psi\right).
  \label{eq:free_fourier}
\end{align}
In the last step of Eq.~\eqref{eq:free_fourier}, we have written $\mathscr{F}_{\text{el}}$ and the quadratic parts of the mixing free energy $f(\psi)$ and the interfacial energy $\frac{1}{2}\kappa\abs{\grad \psi}^{2}$ in Fourier space.
For a Gaussian kernel used in the Letter, $K_{h}(\bm{x}) = (4\pi h^{2})^{-3/2}\mathrm{e}^{-\abs{\bm{x}}^{2}/(4h^{2})}$,
we have $K_{h}(\bm{q}) = \mathrm{e}^{-h^{2}q^{2}}$.
From the above equation, we deduce that the Fourier transform of the effective binary interaction for the $\psi$ field is
\begin{equation}
  F_{\bm{q}} = a(T-\Tc) + \kappa q^{2} + M\text{e}^{-h^{2}q^{2}},
\end{equation}
which completes the derivation of Eq.~\eqref{main:eq:free_fourier} of the Letter.
For computational purposes, it is also useful to consider the real-space representation of Eq.~\eqref{eq:free_fourier}, given by
\begin{equation}
  \mathscr{F}[\psi] = \int \dd^{3}x\,\left[\frac{1}{2}a(T-\Tc)\psi^{2} + \frac{1}{4}b\psi^{4} + \frac{1}{2}\kappa\abs{\grad \psi}^{2} + \frac{1}{2}M\psi\bar{\psi} - \eta\psi\right],
  \label{eq:free_real}
\end{equation}
where $\bar{\psi}$ is the coarse-grained $\psi$ field, defined using Eq.~\eqref{main:eq:coarse} as
\begin{equation}
  \bar{\psi}(\bm{x}) = \int\mathrm{d}^{3}x'\,K_{h}(\bm{x} - \bm{x}')\, \psi(\bm{x}').
  \label{eq:cgpsi}
\end{equation}

\section{Comparison with experiments}
\label{sec:comparison}

\subsection{Rubber elasticity}

Elastomers are broadly defined as crosslinked polymers that display rubber-like elasticity---a property that primarily arises from changes in the configurational entropy of the chains between the crosslinks in the polymer network~\cite{rubinstein2003,tanaka2011,lodge2020}.
Classical theories of entropic elasticity, however, are often insufficient to explain experimental data, and several theories, with varying levels of sophistication, have been devised to describe rubbery materials~\cite{han1999}.
For instance, the behavior of polymer networks can significantly change if the polymer is crosslinked in the melt state, which can introduce entanglement effects that cannot be described using classical theories.
In comparison, classical theories are often sufficient to describe polymer networks crosslinked in semi-dilute solutions~\cite{horkay2023}.
PDMS elastomers, such as the ones we consider in this work, can be particularly susceptible to chain entanglement as they are typically crosslinked in the melt state.

To understand the effect of entanglement interactions, we consider the localization model introduced and refined by Douglas, Gaylord, Horkay, and coworkers~\cite{gaylord1987,gaylord1990,douglas2010,douglas2024,han1999,douglas1993}.
In this model, the nonlinear hyperelastic energy density $\mathscr{W}$, which is a function of the principal stretches $\alpha_{1}$, $\alpha_{2}$, and $\alpha_{3}$, is taken to be of the form
\begin{equation}
  \mathscr{W}(\alpha_{1}, \alpha_{2}, \alpha_{3}) = \underbrace{\frac{1}{2}\left(1 - \frac{2}{\vartheta}\right)\nu\kbt\left(\alpha_{1}^{2} + \alpha_{2}^{2} + \alpha_{3}^{3} - 3\right)}_{\text{network elasticity}}
  + \underbrace{\left[G_{\text{N}} + \left(1-\frac{2}{\vartheta}\right)\omega\nu\kbt\right]\left(\alpha_{1} + \alpha_{2} + \alpha_{3} - 3\right)}_{\text{entanglement effects}}.
  \label{eq:rubber_energy}
\end{equation}
Here $\nu$ is the \emph{strand} density, defined as the average number of strands per unit volume.
A strand is a segment of the polymer network between two adjacent crosslinks, with no intervening crosslinks in between.
Also, $\vartheta$ is the functionality (branch number) of the polymer network.
The first term in Eq.~\eqref{eq:rubber_energy} is the usual free energy density
for network elasticity that is proportional to the strand density $\nu$.
The second term captures the free energy changes due to chain entanglement.
In this term, $G_{\text{N}}$ is the plateau modulus of the polymer melt, which can be interpreted as the shear modulus of the polymer in the near absence of crosslinks (i.e., when $\nu \to 0$).
Additionally, there is also an elastic energy contribution due to entanglements that scales in proportion to the strand density $\nu$ with a material-dependent prefactor $\omega$.

Assuming an incompressible dry (unswollen) polymer network under a uniaxial stretch with $\alpha_{1} = \alpha$ and $\alpha_{2} = \alpha_{3} = \alpha^{-1/2}$, the tensile stress $\sigma_{\text{t}}$, which is what one measures in indentation tests, can be obtained from Eq.~\eqref{eq:rubber_energy} as~\cite{lodge2020}
\begin{equation}
  \sigma_{\text{t}} = \alpha\frac{\partial\mathscr{W}}{\partial \alpha}
  = \left[3\left(1-\frac{2}{\vartheta}\right)\nu\kbt\left(1 + \frac{1}{2}\omega\right) + \frac{3}{2}G_{\text{N}}\right](\alpha - 1) + \mathcal{O}[(\alpha - 1)^{2}].
\end{equation}
Noting that the linear strain $\approx \alpha - 1$, we can read off the (dry) Young's modulus of the elastomer to be
\begin{equation}
  Y = 3\left(1-\frac{2}{\vartheta}\right)\nu\kbt\left(1 + \frac{1}{2}\omega\right) + \frac{3}{2}G_{\text{N}}.
  \label{eq:Y_rubber}
\end{equation}

\begin{figure}
  \begin{center}
    \includegraphics{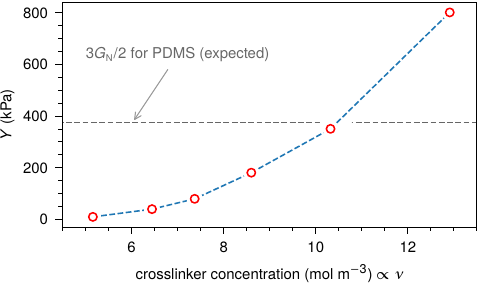}
  \end{center}
  \caption{Young's modulus $Y$ of the elastomers considered in Ref.~\cite{fernandez-rico2024} measured in indentation tests as a function of the crosslinker concentration.
      The strand density $\nu$ and the crosslinker concentration have a monotonic relationship, and we see no visible plateau modulus $G_\text{N}$ as $\nu \to 0$, showing that entanglement contribution to network elasticity is minimal.
  }
  \label{fig:crosslink}
\end{figure}

We expect the number of crosslinks to monotonically increase with the crosslinker concentration, which we take to be a proxy for the strand density $\nu$.
This assumption, for instance, was used to estimate $\nu$ for natural rubber based on the premise that each crosslinker molecule decomposes to form one tetrafunctional crosslink~\cite{mckenna1991,douglas1993}.
The value of the plateau modulus $G_{\text{N}}$ can be estimated from Eq.~\eqref{eq:Y_rubber} via indentation tests in the limit $\nu \to 0$.
From previous studies on polymer networks crosslinked in the melt, $G_{\text{N}}$ values are expected to be a few hundred kPa, e.g., for natural rubber $G_{\text{N}} \sim 560\unit\text{kPa}$~\cite{douglas1993} and for PDMS, $G_{\text{N}} \sim 240\text{--}250\unit\text{kPa}$~\cite{douglas2024}.

Figure~\ref{fig:crosslink} shows the measured Young's moduli for the elastomers in Ref.~\cite{fernandez-rico2024} as a function of the crosslinker concentration.
Remarkably, the data indicates no significant nonzero intercept for small $\nu$.
Additionally, the smallest measured value of $Y = 10\unit\text{kPa}$, is substantially lower than the expected range for $G_{\text{N}}$.
This strongly suggests that entanglement effects are negligible in these elastomers.
Given the absence of more detailed and systematic indentation tests, at the moment, we can only attribute the lack of a measurable $G_{\text{N}}$ to the specific crosslinking technique used in Ref.~\cite{fernandez-rico2024}.
In the absence of experimental estimates for the parameter $\omega$ in Eq.~\eqref{eq:rubber_energy}, the choice $\omega = 1/3$ has been suggested~\cite{douglas2010}.
However, such a choice will not alter our numerical results significantly and because of the minimal evidence for entanglement effects, from here on, we take both $G_{\text{N}}$ and $\omega$ to be zero.
This leads to the classical James--Guth ``phantom'' model of network elasticity~\cite{flory1976,james1944,tanaka2011}, with the Young's modulus of the elastomer in the dry state being
\begin{equation}
  Y = 3\left(1-\frac{2}{\vartheta}\right)\nu\kbt
  = 3\left(1-\frac{2}{\vartheta}\right)\frac{\rho\kbt}{\ms}.
  \label{eq:youngs_dry}
\end{equation}
Here we have expressed the strand density $\nu$ in terms of the average molecular mass $\ms$ of the strands and the dry mass density of the elastomer $\rho$ as $\nu = \rho/\ms$~\cite{lodge2020}.
The above equation is frequently used to experimentally estimate $\ms$ by measuring $Y$~\cite{reinitz2015}.

The experimental results in Ref.~\cite{fernandez-rico2024} are discussed in terms of the Young's modulus of the dry elastomer, given by Eq.~\eqref{eq:youngs_dry}.
However, the rescaled longitudinal modulus $M$ that enters the free energy in Eq.~\eqref{eq:free_real} is that of the swollen elastomer.
Swelling can change the elastic moduli of elastomers significantly, with the moduli acquiring a dependence on the polymer network volume fraction $\phi$~\cite{douglas2024}.
We assume that the elastomer swells isotropically from a dry state composed entirely of the polymer network.
The elastic response of a swollen polymer network is also crucially affected by chain entanglement.
As we judge chain entanglement to be minimal for the elastomers we consider, we shall continue to use results from classical rubber elasticity.
By linearizing the Cauchy stress derived from Eq.~\eqref{eq:rubber_energy} for infinitesimal displacements superimposed on top of the initial swelling, we find the associated Lam\'{e} parameters%
\footnote{The Lam\'{e} parameter $\lambda$ being negative in Eq.~\eqref{eq:moduli} is not a cause for concern because the overall stability of a swollen elastomer is determined by both thermodynamic and elastic contributions to the free energy.
Indeed, this is why it is sometimes customary to work with osmotic moduli~\cite{onuki1989b,onuki1993,doi2009,jia2021}, which take into account both contributions. See Sec.~\ref{sec:fluctuations} for more details.}
$\lambda$ and $\mu$ to be~\cite{onuki1993,jia2021}
\begin{equation}
  \lambda = -\left(1-\frac{2}{\vartheta}\right)\nu\kbt\phi^{1/3}
  \quad\text{and}\quad
  \mu = \left(1-\frac{2}{\vartheta}\right)\nu\kbt\phi^{1/3}.
  \label{eq:moduli}
\end{equation}

We can estimate the polymer volume fraction $\phi$ from the reported mass fraction $c_{\text{oil}}$ of the oil (solvent) present in the swollen elastomer.
Assuming that the mass densities of the elastomer and the oil ($\rho$ and $\rho_{\text{oil}}$, respectively) do not change considerably during the swelling process, the network volume fraction $\phi$ is estimated from $c_{\text{oil}}$ using
\begin{equation}
  \phi = \frac{(1-c_{\text{oil}})\rho_{\text{oil}}}{(1-c_{\text{oil}})\rho_{\text{oil}} + c_{\text{oil}}\rho}.
\end{equation}
The amount of oil absorbed depends on the dry Young's modulus of the elastomer $Y$ and the temperature $T$.
Because of this, the mass fraction $c_{\text{oil}}$ reported in Ref.~\cite{fernandez-rico2024} varies between $c_{\text{oil}} \approx 0.35$ ($Y = 800\unit\text{kPa}$, $T=23\unit^{\circ}\text{C}$) and $c_{\text{oil}} \approx 0.80$ ($Y = 10\unit\text{kPa}$, $T=80\unit^{\circ}\text{C}$).
The density values listed in Table~\ref{tab:parameters}, which also lists other physical parameters, gives us a $\phi$ roughly in the range 0.2--0.7.
In this range, the moduli $\lambda$ and $\mu$ only have a weak dependence on $\phi$, and given that the system is close to the critical point, we take $\phi \approx \phic$ for simplicity in all our estimates.
Using Eqs.~\eqref{eq:youngs_dry} and \eqref{eq:moduli}, we can finally express the rescaled longitudinal modulus $M$ in terms of the dry Young's modulus as
\begin{equation}
  M = \phic^{-2}(\lambda + 2\mu) = \frac{1}{3}\phic^{-5/3}Y.
  \label{eq:longitudinal}
\end{equation}

\subsection{Choice of the length scale \emph{h} and predicted domain size}

In elastomers and gels, the end-to-end distance of the strands within the polymer network is typically used as the characteristic size of the network~\cite{parrish2017,richbourg2020}.
Assuming that the strands behave like a freely-jointed chain modified by a Flory characteristic ratio $C_{\infty}$,
the root-mean-square end-to-end distance $\xi$ before swelling is given by the usual expression~\cite{rubinstein2003,tanaka2011,lodge2020}
\begin{equation}
  \xi^{2} \sim \left(1-\frac{2}{\vartheta}\right)C_{\infty}N\ell^{2}.
  \label{eq:mesh_size_basic}
\end{equation}
Here, $N$ is the degree of polymerization, i.e., the number of repeat units in the strand, and $\ell$ is the length of the
PDMS repeat unit [--$\text{Si}(\text{CH}_{3})_{2}\text{O}$--], taken to be twice the Si--O bond length of $1.64\unit\text{\r{A}}$~\cite{mark2004} in siloxane backbones.
For strands of molecular mass $\ms$, consisting of repeat units
of mass $\mr$, the degree of polymerization is $N = \ms/\mr$~\cite{lodge2020}.
The additional factor of $(1 - 2/\vartheta)$ in Eq.~\eqref{eq:mesh_size_basic} accounts for the effect of junction fluctuations in the phantom model of rubber elasticity~\cite{tanaka2011}.
From here on, we assume the junctions in the polymer network to have perfect tetrafunctional connectivity and set $\vartheta = 4$.
Then, upon using Eq.~\eqref{eq:youngs_dry} to eliminate the strand mass $\ms$ and express $N$ in terms of the Young's modulus $Y$ of the dry elastomer, one finds
\begin{equation}
  N = \frac{\ms}{\mr} = \frac{3\rho\,\kbt}{2Y\mr}.
  \label{eq:repeat_units}
\end{equation}
Using the above equation, we see that for the PDMS elastomer used in the experiments, $N$ ranges from about $60$ (at $Y=800\unit\text{kPa}$) to about $5000$ (at $Y=10\unit\text{kPa}$).
Thus, the end-to-end distance $\xi$ between the crosslinks of an elastomer scales as~\cite{yoo2006,parrish2017}
\begin{equation}
  \xi \sim \left(\frac{3B}{Y}\right)^{1/2}
  \enspace\text{with}\quad B = \frac{C_{\infty}\rho\ell^{2}\kbt}{4\mr}.
  \label{eq:mesh_size}
\end{equation}
The parameter $B$ has dimensions of energy per unit length, and for PDMS we estimate it to be $B = 0.006\unit\text{kPa}\unit\text{\textmu m}^{2}$ (Table~\ref{tab:parameters}).
Equation~\eqref{eq:mesh_size} predicts a $\xi$ roughly between $5\unit\text{nm}$ (at $Y = 800\unit\text{kPa}$) and $50\unit\text{nm}$ (at $Y = 10\unit\text{kPa}$).
Here we note that for swollen elastomers, $\xi$ is sometimes multiplied by a factor of $\phi^{-1/3}$ to account for a change in the distance between the crosslinks due to isotropic swelling of the elastic background (affine assumption)~\cite{canal1989,richbourg2020}.
Even though such a factor will not affect our scaling results, it is not consistent with the physical assumptions of the phantom model of rubber elasticity that we use, where the junction fluctuations are considered to be independent of the deformation.

\begin{table*}
\centering
\caption{Physical and model parameters.}
\label{tab:param}
\begin{ruledtabular}
\begin{tabular}{l l l l}
  Parameter & Description & Value\\
  \hline
  $\kbt$ & Thermal energy at $T = 300\unit\text{K}$ & $4.14\times 10^{-21}\unit\text{J}$\\
  $\mr$ & Molecular mass of PDMS repeat unit [--$\text{Si}(\text{CH}_{3})_{2}\text{O}$--] & $1.23\times10^{-25}\unit\text{kg}$ ($74.2\unit\text{g}\unit\text{mol}^{-1}$)\\
  $\ell$ & Length of the PDMS repeat unit~\cite{mark2004} & $3.28\unit\text{\r{A}}$\\
$\rho$ & Mass density of PDMS\footnote{Vinyl terminated polydimethylsiloxane (\href{https://www.gelest.com/product/DMS-V31}{DMS-V31})
safety data sheet (Gelest, Morrisville, PA, 2014).} & $970\unit\text{kg}\unit\text{m}^{-3}$\\
$\rho_{\text{oil}}$ & Mass density of the solvent (heptafluorobutyl methacrylate\footnote{2,2,3,3,4,4,4-Heptafluorobutyl methacrylate
(\href{https://store.apolloscientific.co.uk/product/1h1h-perfluorobutyl-methacrylate}{PC11102}) safety data sheet (Apollo Scientific, Bredbury, UK, 2023).}) & $1345\unit\text{kg}\unit\text{m}^{-3}$\\
  $C_{\infty}$ & Flory characteristic ratio for PDMS~\cite{rubinstein2003} & 6.8\\
  $\kappa$ & Interfacial parameter ($\sim \kbt/\ell$~\cite{leibler1987}) & $0.013\unit\text{kPa}\unit\text{\textmu m}^{2}$ ($1.3\times10^{-11}\unit\text{J}\unit\text{m}^{-1}$)\\
  $B$ & $C_{\infty}\rho\ell^{2}\kbt/(4\mr)$ [see Eq.~\eqref{eq:mesh_size}] & $0.006\unit\text{kPa}\unit\text{\textmu m}^{2}$ ($6.0\times10^{-12}\unit\text{J}\unit\text{m}^{-1}$)\\
  $\Tc$ & Critical temperature in the Landau free energy $f(\psi)$ [Eq.~\eqref{main:eq:landau}] & $70\unit{^\circ}\text{C}$\\
  $\phic$ & Critical volume fraction in $f(\psi)$ & 0.2\\
  $a$ & Quadratic coefficient in $f(\psi)$ & $0.025\unit\text{kPa}\unit\text{K}^{-1}$\\
  $b$ & Quartic coefficient in $f(\psi)$ & $2\unit\text{kPa}\unit\text{K}^{-1}$\\
  $n$ & Average number of crosslinks coarse-grained over & 110
\end{tabular}
\label{tab:parameters}
\end{ruledtabular}
\end{table*}

As we have remarked in the Letter, the coarse-graining length scale is $h = n\xi$, with $n$ being the average number of crosslinks over which we coarse-grain in each direction.
Using Eqs.~\eqref{eq:mesh_size} and~\eqref{eq:longitudinal} we find the dimensionless parameter $\gamma$ to be
\begin{equation}
  \gamma = Mh^{2}\kappa^{-1} = Bn^{2}\kappa^{-1}\phic^{-5/3}.
  \label{eq:gamma}
\end{equation}
Note that $\gamma$ is independent of the Young's modulus $Y$, and it satisfies the condition $\gamma > 1$ for the parameter values in Table~\ref{tab:param}.
We now see that the domain size $\Lambda$ scales as
\begin{equation}
  \Lambda \sim 2\pi\qm^{-1} = 2\pi h(\ln\gamma)^{-1/2} = 2\pi\left[\frac{3Bn^{2}}{Y\ln\left(Bn^{2}\kappa^{-1}\phic^{-5/3}\right)}\right]^{1/2}
  = 2\pi\left[N\frac{C_{\infty}n^{2}\ell^{2}}{\ln\left(Bn^{2}\kappa^{-1}\phic^{-5/3}\right)}\right]^{1/2}\hspace{-1em}.
  \label{eq:domain_size}
\end{equation}
This completes the derivation of Eq.~\eqref{main:eq:domain_size} of the Letter.
In the last step above, using Eq.~\eqref{eq:repeat_units}, we have expressed $\Lambda$ in terms of the number $N$ of repeat units between the crosslinks of the polymer network.
A similar result was predicted by \citet{gennes1979} for the microphase domain size in crosslinked polymer blends using a phenomenological model different from our nonlocal model.

\begin{figure}
  \begin{center}
    \includegraphics{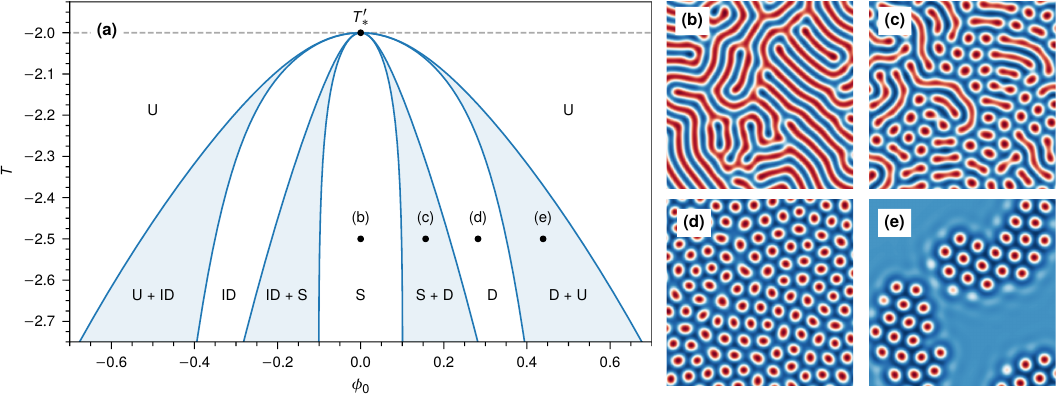}
  \end{center}
  \caption{%
    (a) General phase diagram in the $(\phiz,T)$ plane obtained by choosing general, unitless parameters $\Tc = \phic = 0$, $a = b = \kappa = h = 1$, and $M = \mathrm{e}$ (Euler's number).
    With these choices, the order parameter $\psi = \phi$, $\gamma = Mh^{2}/\kappa = \mathrm{e}$, and $\qm^{2} = h^{-2}\ln\gamma = 1$, giving rise to domains of size $\Lambda \sim 2\pi$.
    Above the temperature $\Tc' = \Tm(0) = -2$ only the uniform (U) phase is stable.
    Microphase separation occurs below $\Tc'$ and leads to the formation of phases composed of stripes (S), droplets (D), and inverted droplets (ID).
    Close to $\Tc'$, there are also four regions of phase coexistence (shaded in blue) between the uniform phase and inverted droplets (U + ID), between inverted droplets and stripes (ID + S), between stripes and droplets (S + D), and between droplets and the uniform phase (D + U).
    (b)--(e) Equilibrium profiles at $T = -2.5$ for various values of the mean order parameter $\phiz$ (marked in the phase diagram).
    Solvent-rich ($\phi < 0$) and solvent-deficient ($\phi > 0$) regions are highlighted in red and blue, respectively.
  The equilibrium profiles were obtained by numerically minimizing the free energy (Sec.~\ref{sec:numerical}) in a square domain of size 36$\times$36.
  }
  \label{fig:general_phase_diagram}
\end{figure}

\section{Phase diagram}
\label{sec:phase}

We can use the one-mode approximation to analytically construct the phase diagram in the weak-segregation limit.
Although the free energy in Eq.~\eqref{main:eq:free_fourier} of the Letter is defined over 3D space, we restrict ourselves to an analysis of 2D modulations for simplicity.
In 2D, one typically considers two modulated phases: the stripe and droplet (hexagonal) phases and the uniform phase, devoid of modulations.
Once the free energy for each of these phases is determined, the phase diagram can be determined via Maxwell construction.
For brevity in presenting the results below, we define a dimensionless parameter
\begin{equation}
  \Gamma = b^{-1}\left[a(T-\Tc) + 3b\psiz^{2} + M \gamma^{-1}(1 + \ln\gamma)\right].
  \label{eq:gamma_param}
\end{equation}

\paragraph*{Uniform phase.}  The uniform phase has a free energy density
\begin{equation}
  f_{\text{U}} = \frac{1}{2}\left[a(T-\Tc) + M\right]\psiz^{2} + \frac{1}{4}b\psiz^{4}.
\end{equation}

\paragraph*{Stripe phase.}  For the stripe phase, we consider the stripe solution
$\psi_{\text{S}} = \psiz + A\cos(qx)$, with $A\cos(qx)$ representing a modulation of amplitude $A$ and wavenumber $q$ directed along one of the spatial directions.
We substitute this solution in Eq.~\eqref{eq:free_real} and minimize the free energy with respect to $q$ and $A$
to obtain the free-energy density $f_{\text{S}}$ of the stripe phase at $q = \qm$ and $A = \Am$ as
\begin{equation}
  f_{\text{S}} = f_{\text{U}} - \frac{b}{6}\Gamma^{2},
  \quad\text{with}\quad
  \qm^{2} = h^{-2}\ln\gamma
  \quad\text{and}\quad
  \Am^{2} = -\frac{4}{3}\Gamma.
\end{equation}
As the modulation amplitude must be real for the stripe solution to exit, setting $\Gamma = 0$ in Eq.~\eqref{eq:gamma_param} gives an estimate for the microphase separation temperature $\Tm$, which we find to be
\begin{equation}
  \Tm = \Tc - a^{-1}\left[3b\psiz^{2} + M \gamma^{-1}\left(1 + \ln \gamma\right)\right].
\end{equation}
The same expression for $\Tm$ can also be derived by linear stability analysis of the fluctuations around the uniform phase, as in Eq.~\eqref{main:eq:tm} of the Letter.

\paragraph*{Droplet phase.}  For the droplet (hexagonal) phase, we consider the solution~\cite{elder2004}
\begin{equation}
  \psi_{\text{D}} = \psiz + A\left[\cos(qx)\cos(qy/\sqrt{3}) - \frac{1}{2}\cos(2qy/\sqrt{3})\right].
\end{equation}
As with the stripe phase, we use this solution in Eq.~\eqref{eq:free_real} and minimize the free energy with respect to $q$ and $A$ to find
\begin{equation}
  \qm^{2} = h^{-2}\ln\gamma
  \quad\text{and}\quad
  A_{\text{m},\pm} = \frac{4}{5}\left[\psiz \pm \frac{1}{3}\left(9\psiz^{2} - 15\Gamma\right)^{1/2}\right].
\end{equation}
The solution $A_{\text{m},+}$ corresponds to the droplet phase, and $A_{\text{m},-}$ corresponds to the ``inverted'' droplet phase,
with the minimized free-energy density given by
\begin{equation}
  f_{\text{D}} = f_{\text{U}} - \frac{3b}{64}A_{\text{m},\pm}^{2}\left(\psiz A_{\text{m},\pm} - 2\Gamma\right).
\end{equation}

We use Maxwell construction to equate each phase's chemical potential and osmotic pressure to find the coexistence curves between the different phases.
This leads to the following set of equations, which, when solved numerically, gives the mean value of the order parameter in each of the phases:
\begin{equation}
  \frac{\partial f_{i}}{\partial \psi_{0,i}} = \frac{\partial f_{j}}{\partial \psi_{0,j}}
  \quad\text{and}\quad
  \psi_{0,i}\left(\frac{\partial f_{i}}{\partial \psi_{0,i}}\right) - f_{i}
  =
  \psi_{0,j}\left(\frac{\partial f_{j}}{\partial \psi_{0,j}}\right) - f_{j}.
\end{equation}
Here, the subscripts $i, j$ refer to one of the three phases: U, S, or D.

A phase diagram constructed using the free-energy densities derived above for 2D modulations and choosing general, unitless parameters is presented in Fig.~\ref{fig:general_phase_diagram}(a).
Close to the critical point, three distinct phases can be observed: a uniform phase, a droplet (hexagonal) phase characterized by solvent-rich droplets dispersed within a solvent-deficient region, and a stripe phase consisting of alternating solvent-rich and solvent-deficient layers.
Additionally, there is an inverted droplet phase consisting of solvent-deficient ``droplets'' embedded in a solvent-rich region.
Compared to the phase diagram in Fig.~\ref{main:fig:phase} of the Letter, here we see the four phase coexistence regions more conspicuously.
The general predictions of the phase diagram agree rather well with numerically obtained equilibrium morphologies in Figs.~\ref{fig:general_phase_diagram}(b)--\ref{fig:general_phase_diagram}(e).
In Fig.~\ref{fig:general_phase_diagram}(a), all phase transitions are first order except at the critical point where a continuous transition from the uniform phase to the stripe phase is possible.

\begin{figure}
  \begin{center}
    \includegraphics{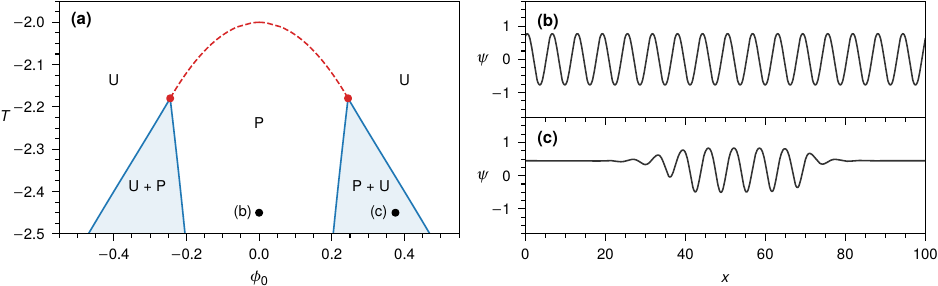}
  \end{center}
  \caption{(a) General phase diagram in the $(\phiz, T)$ plane considering only 1D modulations for the same parameters as in Fig.~\ref{fig:general_phase_diagram}.
      The dashed red curve indicates a second-order transition line, across which a continuous phase transition between the uniform (U) and the patterned (P) phases can occur.
      Solid blue curves are first-order transition lines.
      These curves meet at two tricritical points (marked with red dots).
      Regions of phase coexistence between the patterned and the uniform phases (P + U) are shaded in blue.
      (b),(c) Example equilibrium profiles $\psi(x)$ for various values of the mean order parameter $\psiz$ (marked in the phase diagram).
  }
  \label{fig:phase_1d}
\end{figure}

Intriguingly, the phase behavior of the system in 1D (Fig.~\ref{fig:phase_1d}) is somewhat distinct from the 2D situation.
In 1D, the only modulated phase that one considers is a ``patterned'' phase $\psi(x) = \psiz + A\cos(qx)$, equivalent to the stripe phase in 2D and 3D~\cite{villain-guillot1998}.
From Fig.~\ref{fig:phase_1d}, we see that in the temperature range immediately below the critical point, the system can transition from the uniform phase to the patterned phase continuously (indicated by a second-order transition curve.%
\footnote{Similar second-order transition curves have also been predicted in the 1D version of the phase-field crystal model~\cite{elder2004,thiele2019} and in phenomenological models of membrane adhesion~\cite{komura2000} analogous to the Blume--Emery--Griffiths model for $^{3}$He-$^{4}$He mixtures~\cite{chaikin1995}.})
Phase separation becomes first order below a tricritical point, leading to the emergence of a coexistence region between the uniform and patterned phases.
The second-order transition curve in Fig.~\ref{fig:phase_1d}, given by $\Gamma = 0$ [see Eq.~\eqref{eq:gamma_param}], also exists in the 2D phase diagram of Fig.~\ref{fig:general_phase_diagram}(a).
However, it always falls within the phase boundaries of the other phases and for this reason, a continuous transition to the patterned (stripe) phase is not generally seen in 2D or 3D.
Here it is worth noting that the recent study by \citet{qiang2024} on the strong-segregation limit of microphase separation in elastomers~\cite{qiang2024} also identifies a second-order phase transition at high stiffnesses and the presence of tricritical points.
While their model differs from ours in several key aspects, the fact that their analysis is in 1D is the likely explanation for the observed continuous transition to patterned phases.

\subsection*{Similarity to other pattern-forming systems}

The phase diagram in Fig.~\ref{fig:general_phase_diagram} is rather similar to phase diagrams of other pattern-forming systems such as block copolymers and Langmuir monolayers~\cite{seul1995}.
Given this similarity it is worth examining the connection further.
If we are primarily interested in modulations in $\psi$ with wavelengths much larger than the coarse-graining length scale, i.e., when $h^{2}q^{2} \ll 1$, we can expand $\bar{\psi}_{\bm{q}} = K_{h}(\bm{q})\psi_{\bm{q}} = \mathrm{e}^{-h^{2}q^{2}}\psi_{\bm{q}}$ in Eq.~\eqref{eq:free_fourier} in powers of $q$ to find
\begin{equation}
  \bar{\psi}(\bm{q}) = \left[1 - h^{2}q^{2} + \frac{1}{2}h^{4}q^{4} + \cdots\right]\psi(\bm{q}).
\end{equation}
Taking the inverse Fourier transform of the above expression, we identify the real-space representation of the coarse-grained order parameter $\bar{\psi}(\bm{x})$ to be
\begin{equation}
  \bar{\psi}(\bm{x}) = \left[1 + h^{2}\nabla^{2} + \frac{1}{2}h ^{4}\nabla^{4} + \cdots\right] \psi(\bm{x}).
\end{equation}
Substituting this in Eq.~\eqref{eq:free_real} and after sufficient integration by parts to ensure that only powers of $\psi$ and $\nabla^{2}\psi$ remain, we find that the total free energy takes the form
\begin{equation}
  \mathscr{F}[\psi] \sim \int\mathrm{d}^{3}x\,\left\{\frac{1}{2}a(T-T_{**})\psi^{2} + \frac{1}{4}b\psi^{4} + \frac{1}{4}Mh ^{4}\left[(\nabla^{2} + \qm^{2})\psi\right]^{2} - \eta\psi\right\},
  \label{eq:free_expand}
\end{equation}
where the wavenumber $\qm$ and the temperature $T_{**}$ are given by
\begin{equation}
  \qm^{2} = h ^{-2}(1-\gamma^{-1})
  \quad\text{and}\quad
  T_{**} = \Tc - a^{-1}M\left[1-\frac{1}{2}(1-\gamma^{-1})^{2}\right].
\end{equation}
The above values agree to $\mathcal{O}[(\gamma - 1)^{2}]$ with the corresponding expressions in Eq.~\eqref{main:eq:qm} and Eq.~\eqref{main:eq:tm} of the Letter in the limit $\gamma \to 1^{+}$.
The free energy in Eq.~\eqref{eq:free_expand} is the Landau--Brazovskii free energy, which has been used to model a host of pattern-forming systems.
Within the context of block copolymers, Fredrickson and Helfand~\cite{fredrickson1987} simplified Leibler's free energy functional~\cite{leibler1980} for diblock copolymers to the above form.
Apart from the phases in Fig.~\ref{fig:general_phase_diagram}, a wide range of equilibrium morphologies, including BCC and gyroid phases, have been predicted for systems following this free energy in 3D~\cite{yamada2008}.

\section{Fluctuation effects, scattering, and Lifshitz behavior}
\label{sec:fluctuations}

To understand the effect of fluctuations above the transition temperature, we set $\psi = \psiz + \delta\psi$ and expand Eq.~\eqref{eq:free_fourier} to $\mathcal{O}(\delta\psi^{2})$, and find the quadratic (Gaussian) free energy of the fluctuations $\delta\psi$ in Fourier space as
\begin{equation}
  \mathscr{F}_{\text{G}}[\delta\psi] = \frac{1}{2} \int \frac{\dd^{3}q}{(2\pi)^{3}}(F_{\bm{q}} + 3b\psiz^{2})\,\delta\psi_{\bm{q}}\,\delta\psi_{-\bm{q}}.
\end{equation}
The intensity distribution due to fluctuations seen in scattering experiments at equilibrium is proportional to the static structure factor
$S(\bm{q})= \langle\delta\psi_{\bm{q}}\,\delta\psi_{-\bm{q}}\rangle$.
For a quadratic free energy functional such as $\mathscr{F}_{\text{G}}$, it is given by~\cite{chaikin1995}
\begin{align}
  S(\bm{q}) = \frac{\int \mathcal{D}[\delta\psi_{\bm{q}}]\,\mathcal{D}[\delta\psi_{-\bm{q}}]\,\delta\psi_{\bm{q}}\,\delta\psi_{-\bm{q}}\,\mathrm{e}^{-\mathscr{F}_{\text{G}}/(\kbt)}}
         {\int \mathcal{D}[\delta\psi_{\bm{q}}]\,\mathcal{D}[\delta\psi_{-\bm{q}}]\,\mathrm{e}^{-\mathscr{F}_{\text{G}}/(\kbt)}}
  &= \frac{\kbt}{F_{\bm{q}} + 3b\psiz^{2}}\\
            &= \frac{\kbt}{a(T-\Tc) + 3b\psiz^{2} + \kappa q^{2} + M\mathrm{e}^{-h^{2}q^{2}}}.
  \label{eq:sfactor}
\end{align}
For $M=0$, the structure factor has an Ornstein--Zernike form $S(q) \sim [1 + (qr)^{2}]^{-1}$ with a mean-field correlation length $r \sim (T-\Tc)^{-1/2}$ as expected in general phase-separating systems.
When $M = \phic^{-2}(\lambda + 2\mu) \neq 0$ and $h = 0$, i.e., when there is no mesoscopic length scale associated with elasticity, $S(\bm{q})$ would have the same form, but with a shifted critical temperature.
The associated thermal correlation length close to the critical point is given by
\begin{equation}
  r^{2} = \frac{\kappa \phic^{2}}{a(T-\Tc)\phic^{2} + \lambda + 2\mu} = \frac{\kappa \phic^{2}}{K_{\text{os}} + 4\mu/3}.
  \label{eq:corr_length}
\end{equation}
In the last expression, we have written the denominator in terms of the osmotic bulk modulus $K_{\text{os}} = \phic^{2}f''(\phic) + \lambda + 2\mu/3$ noting that $f''(\phic) = a(T-\Tc)$.
In general, isotropic elastic systems become mechanically unstable when $K_{\text{os}} < 0$.
However, because the shear modulus $\mu$ is nonzero, we see that the correlation length $r$ in Eq.~\eqref{eq:corr_length} diverges only when $K_{\text{os}} \to -4\mu/3$.
This is a well known result, particularly in the context of gels~\cite{onuki1993} where a distinction is made between the mechanical instability point given by $K_{\text{os}} = 0$, and the ``cloud point'' given by $K_{\text{os}} + 4\mu/3 = 0$ where the gel loses thermodynamic stability.
However, macroscopic shape changes in gels occur over a significant amount of time and the mechanical instability at $K_{\text{os}} = 0$ can be hard to observe unless the gel sample is very small in size~\cite{onuki1988a,onuki1999}.
For this reason, gels can remain stable even when $K_{\text{os}} < 0$ and can exhibit anomalous properties like a negative Poisson's ratio when $K_\text{os} < 2\mu/3$~\cite{hirotsu1991}.
At temperatures below the cloud point, the gel becomes opaque and undergoes spinodal decomposition~\cite{tanaka1977,onuki1989b}.

\begin{figure}
  \begin{center}
    \includegraphics{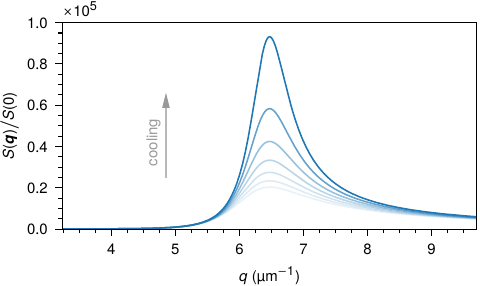}
  \end{center}
  \caption{The structure factor in Eq.~\eqref{eq:sfactor} for an elastomer of Young's modulus $Y = 800\unit\text{kPa}$ and $\psiz = 0.32$ for temperatures $T = 30.0\unit{^{\circ}\text{C}}, \ldots, 24.0\unit{^{\circ}\text{C}}$ (counting from bottom to top).
    The rescaled longitudinal modulus $M \approx 3.9\unit\text{mPa}$ and the length scale $h \approx 0.52\unit\text{{\textmu}m}$, estimated using the results in Sec.~\ref{sec:comparison}.
    Other parameters are the same as in Figs.~\ref{main:fig:comparison} and \ref{main:fig:phase} of the Letter (also in Table~\ref{tab:param}).
  The structure factor peak at $\qm \approx 6.5\unit\text{{\textmu}m}^{-1}$ grows smoothly as the temperature is reduced to $\Tm = 23\unit{^{\circ}\text{C}}$, consistent with both the experimental observation of intensity changes and the final formation of domains of size $\sim 1\unit\text{{\textmu}m}$ at this stiffness~\cite{fernandez-rico2024}.
  }
  \label{fig:sfactor}
\end{figure}

When elasticity operates at a nonzero mesoscopic length scale $h$ and the parameter $\gamma = Mh^{2}/\kappa > 1$, the structure factor has a peak at $\qm^{2} = h^{2}\ln\gamma$.
Furthermore, the peak value $S(\qm) \sim (T - \Tm)^{-1}$, with $\Tm$ being the microphase separation temperature defined in Eq.~\eqref{main:eq:tm} of the Letter.
An example $S(\bm{q})$ for the experimental parameter ranges is illustrated in Fig.~\ref{fig:sfactor}.
When the elastomer is cooled from $T \to \Tm$, we expect the scattering intensity to have a smoothly growing peak at a fixed $q$ similar to $S(\bm{q})$, in agreement with the experiments of Ref.~\cite{fernandez-rico2024}.
In the context of crosslinked polymer blends, de Gennes's prediction of $S(0) = 0$~\cite{gennes1979} was not observed experimentally~\cite{briber1988}, and lead to other theoretical efforts~\cite{read1995} to explain the discrepancy.
Interestingly, our model predicts $S(0) \neq 0$.
This, along with the similarity in the scaling behaviors of the domain size and the transition temperature, strongly suggests that crosslinked polymer blends may also have an internal elastic response that is nonlocal.

To glean more analytical insights into the general nature of the long-range correlations associated with the structure factor in Eq.~\eqref{eq:sfactor},
assuming small $q$, we expand the denominator to $\mathcal{O}(q^{4})$ to find
\begin{equation}
  S(\bm{q}) \sim \frac{S(0)\,\tau}{\tau + 2(1 - \gamma)q^{2} + \gamma h^{2}q^{4}},
  \quad
  \text{where}
  \quad
  \tau = 2\kappa^{-1}[a(T-\Tc) + 3b\psiz^{2} + M].
  \label{eq:sfactor_approx}
\end{equation}
This structure factor can also be derived from the simplified free energy in Eq.~\eqref{eq:free_expand} and resembles the structure factor observed in microemulsions---fluid mixtures of water and oil stabilized by a surfactant~\cite{teubner1987}.
The Fourier inverse of the structure factor $S(\bm{q})$ is the real-space two-point correlation function
$C(\bm{x}) = \langle\delta\psi(\bm{x})\delta\psi(0)\rangle$.
Intuitively, $C(\bm{x})$ is a measure of the probability of the values of $\delta\psi$ at a point $\bm{x}$ and at the origin being the same.

\begin{figure}
  \begin{center}
    \includegraphics{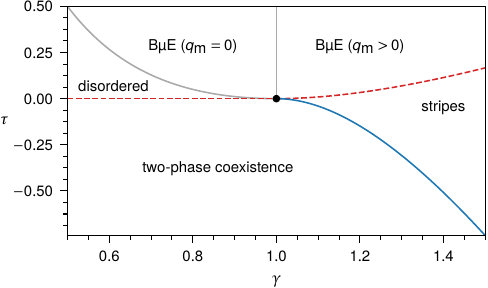}
  \end{center}
  \caption{General phase diagram in the $(\tau, \gamma)$ plane with $h = 1$.
      Here $\tau$ is an effective temperature and $\gamma$ is a dimensionless parameter analogous to the (inverse) elastocapillary number.
      See Eq.~\eqref{eq:sfactor_approx} for definitions.
      The dashed red curve given by $\zeta = -1$ ($\gamma > 1$) and the dashed red line given by $\tau = 0$ ($\gamma < 1$) indicate second-order transitions.
      A first-order triple line (solid blue curve) given by $\tau = -(2+\sqrt{6})(1-\gamma)^{2}/(h^{2}\gamma)$ separates the two-phase coexistence region from the stripe phase.
      The gray curve separating the bicontinuous microemulsion (B{\textmu}E) phase with $\qm = 0$ and the disordered one is the disorder line where $\zeta = 1$.
      The gray vertical line where $\zeta = 0$ $(\gamma = 1)$ represents the Lifshitz line which ends at the Lifshitz point (black dot)~\cite{hornreich1975}.
      Neither the disorder line nor the Lifshitz line indicate phase transitions.}
  \label{fig:lifshitz}
\end{figure}

For clarity in presenting the results below, we define the following nondimensional parameter:
\begin{equation}
  \zeta = \frac{(1 - \gamma)}{\sqrt{\gamma \tau h^{2}}}.
  \label{eq:zeta}
\end{equation}
When $\abs{\zeta} < 1$, the correlation function associated with Eq.~\eqref{eq:sfactor_approx} is of the form~\cite{teubner1987,komura2007}
\begin{equation}
  C(\bm{x}) = C(0)\,\mathrm{e}^{-\abs{\bm{x}}/r}\sinc\left(\frac{2\pi \abs{\bm{x}}}{d}\right),
  \label{eq:correlation}
\end{equation}
and represents exponentially decaying oscillations with a period $d$ and correlation length $r$, given by
\begin{equation}
  r = \left(\frac{\gamma h^{2}}{\tau}\right)^{1/4}\sqrt{\frac{2}{1 + \zeta}}
  \quad
  \text{and}
  \quad
  d = \left(\frac{\gamma h^{2}}{\tau}\right)^{1/4}\sqrt{\frac{2}{1 - \zeta}}.
\end{equation}
In microemulsions, a parameter analogous to $\zeta$ measures the strength of the surfactant: a stronger surfactant results in a negative $\zeta$ and vice versa~\cite{komura2007}.
For elastomers, we see from Eq.~\eqref{eq:zeta} that $\zeta$ is controlled by the parameter $\gamma = Mh^{2}/\kappa$, which is analogous to the (inverse) elastocapillary number, and measures the relative importance of the elastic and surface contributions to the free energy: $\gamma > 1$ results in $\zeta < 0$ and vice versa.
The phase behavior of elastomers above their transition temperature can also be compared to that of microemulsions, as outlined below and graphically represented in Fig.~\ref{fig:lifshitz}.
\begin{itemize}
  \item When $\abs{\zeta} < 1$, both $r$ and $d$ are finite and we consider the system to be in a structured-disordered phase (also called the ``middle'' or the bicontinuous microemulsion {B{\textmu}E} phase), referring to the presence of fluctuating mesoscopic structures within an overall disordered system~\cite{komura2007}.
        This is indicated by a correlation function with decaying oscillations.

  \item When $0 < \zeta < 1$ (which requires $\gamma < 1$), the structure factor has a peak only at $\qm = 0$,
        even though $C(\bm{x})$ continues to have an oscillatory behavior.
        As $\zeta \to 1$, the oscillation period $d$ diverges and the curve where $\zeta = 1$ is the disorder line~\cite{komura2007}.

        For $\zeta > 1$, the correlation function $C(\bm{x})$ decays monotonically without oscillations (disordered phase).
        When $\zeta \gg 1$, the coefficient of $q^{2}$ in the denominator of Eq.~\eqref{eq:sfactor_approx} is large and positive and the $q^{4}$ term can be neglected.
        This leads to an Ornstein--Zernike-like $S(\bm{q})$, with a correlation function that decays exponentially.

  \item The line where $\zeta = 0$ (or $\gamma = 1$) is the Lifshitz line~\cite{hornreich1975}.
        Beyond the Lifshitz line, when $\gamma > 1$ and $-1 < \zeta < 0$, the structure factor has a peak at $\qm^{2} = h^{-2}(1 - \gamma^{-1})$ signifying the dominant presence of fluctuating structures.
        When $\zeta \to -1$, the correlation length $r$ diverges, indicating spatial ordering and the emergence of an ordered phase.
        Microphases appear for $\zeta < -1$.
\end{itemize}

Finally, it should be emphasized that in the presence of fluctuations, the mean-field phase diagrams we have considered accurately describe the phase behavior only when sufficiently far from the critical point.
A more complete analysis of the effect of fluctuations for systems described by the free-energy functional of Eq.~\eqref{eq:free_expand} was presented by Brazovskii using a Hartree approximation~\cite{brazovskii1975}.
Fluctuations were observed to suppress the second-order phase transition at the critical point, which gets replaced by a first-order transition line across which the system can transition from a uniform phase to a stripe phase.
Away from the critical point, fluctuations effectively lead to specific compositional ranges of the mean order parameter $\psiz$ (or $\phiz$) that allow for a direct transition from the uniform phase to various ordered phases as temperature changes.
For block copolymers, fluctuation effects can be significant~\cite{koga1999,komura2008}.
Since the free energy in our model can be simplified to the form in Eq.~\eqref{eq:free_expand}, we anticipate that fluctuations will influence the phase behavior very close to the order-disorder transition temperature.
For example, with the inclusion of fluctuations, a direct transition between the disordered and gyroid phases may occur~\cite{hamley1997}, which could, for example, explain the experimental observation of the gyroid-like bicontinuous phase for elastomers at high stiffnesses~\cite{fernandez-rico2024}.

\section{General coarse-graining procedure}
\label{sec:coarse}

We now describe a general procedure to construct a large class of useful coarse-graining kernels in 3D.
Although this procedure is not the only method, many routinely used kernels can be constructed using this method.

Consider a general probability density function $P(x)$ of a real random variable.
Now, consider the family of densities $P_{h}(x)$, indexed by a parameter $h > 0$, and defined by
\begin{equation}
  P_{h}(x) = h^{-1}P\left(h^{-1}x\right).
\end{equation}
The density $P_{h}(x)$ is also normalized, and it satisfies~\cite{aguirregabiria2002}
\begin{equation}
  \lim_{h\to 0^{+}} P_{h}(x) = \delta(x),
  \label{eq:delta_lim}
\end{equation}
where $\delta(x)$ is the Dirac delta function.
We additionally assume that $P(x)$ has a well defined second moment $\mu_{2} = \int_{0}^{\infty}\dd{x}\,x^{2}P(x)$ on the half-interval $[0, \infty)$.
We can now use the density $P_{h}(x)$ to define a spherically-symmetric convolution kernel $K_{h}(\bm{x})$ in 3D as
\begin{equation}
  K_{h}(\bm{x}) = (4\pi\mu_{2}h^{2})^{-1}P_{h}(\abs{\bm{x}}).
\end{equation}
Using the 3D analogue of Eq.~\eqref{eq:delta_lim}, it can be shown that $\lim_{h\to0^{+}}K_{h}(\bm{x}) =
\delta^{3}(\bm{x})$~\cite{aguirregabiria2002} and we identify the parameter $h$ with the coarse-graining length.
For example, if we choose $P(x)$ to be the Gaussian distribution $P(x) = (4\pi)^{-1/2} \mathrm{e}^{-x^{2}/4}$ with $\mu_{2} = 1$,
we have $K_{h}(\bm{x}) = (4\pi h^{2})^{-3/2} \mathrm{e}^{-|\bm{x}|^{2}/(4h^{2})}$ as in the Letter.

Because $K_{h}(\bm{x})$ is spherically symmetric by construction,
it is natural to compute its Fourier transform $K_{h}(\bm{q})$ in
spherical polar coordinates.
Without loss of generality, we take the polar axis along $\bm{q}$ so that $\bm{q}\cdot\bm{r} = qr\cos{\theta}$, where $r$ is the radial coordinate and $q = \abs{\bm{q}}$.
After integrating over the polar angle $\theta$ as well as the azimuthal angle, we find
\begin{align}
  K_{h}(\bm{q}) &= (\mu_{2}h^{2})^{-1}\int_{0}^{\infty} \mathrm{d}r\,r^{2}\sinc(qr)\,P_{h}(r)\NN\\
                  &= \mu_{2}^{-1}\int_{0}^{\infty} \mathrm{d}r\,r^{2}\sinc(h qr)\,P(r) = K_{1}(h q),
                  \label{eq:kernel_fourier}
\end{align}
showing that $K_{h}(\bm{q})$ is a function%
\footnote{For a general kernel $K_{h}(\bm{x})$, this can also be seen using dimensional arguments.
As $K_{h}(\bm{x})$ is normalized to unity, it must have dimensions of inverse volume, making its Fourier transform $K_{h}(\bm{q})$ dimensionless.
Because the wavenumber $q$ has dimensions of inverse length, and $h$ is the only other length scale that appears in its definition, $K_{h}(\bm{q})$ can only be a function of $hq$.}
of the product $hq$.%

The Fourier transform of the effective binary interaction in Eq.~\eqref{eq:free_fourier} is
\begin{equation}
  F_{\bm{q}} = a(T-\Tc) + \kappa q^{2} + MK_{h}(\bm{q}).
  \label{eq:general_binary}
\end{equation}
The wavenumber $\qm$ is where $F_{\bm{q}}$ acquires its minimum.
Using Eq.~\eqref{eq:kernel_fourier}, we find that $\qm$ satisfies the equation
\begin{equation}
  2\kappa \qm + MK_{h}'(\qm) = 2\kappa \qm + Mh K_{1}'(h \qm) = 0.
\end{equation}
Defining the parameter $\gamma = Mh^{2}/\kappa$, we can write the above equation as
\begin{equation}
  2 + \gamma\left[(h \qm)^{-1}K_{1}'(h \qm)\right] = 0.
  \label{eq:qm}
\end{equation}
The coarse-graining length scale $h \sim \xi \sim Y^{-1/2}$ using Eq.~\eqref{eq:mesh_size} and the rescaled longitudinal modulus $M \sim Y$ from Eq.~\eqref{eq:longitudinal}.
Because of this, we see that the parameter $\gamma$ is independent of both the Young's modulus $Y$ and $h$, and Eq.~\eqref{eq:qm} would generally be a transcendental equation in the product $h \qm$.
Any valid solution to this equation must always scale as $\qm \sim h^{-1} \sim Y^{1/2}$ and the product $h\qm$ would be a function of $\gamma$ alone.
Therefore, irrespective of the kernel $K_{h}$, we see that the domain size $\Lambda$ follows the scaling
\begin{equation}
  \Lambda \sim 2\pi \qm^{-1} \sim Y^{-1/2}.
\end{equation}

The microphase separation temperature $\Tm$ is the temperature at which the fluctuations around the uniform state become linearly unstable.
As this occurs when $F_{\bm{q}} = -3b\psiz^{2}$ and $q = \qm$, using Eqs.~\eqref{eq:kernel_fourier} and~\eqref{eq:general_binary}, we find
\begin{equation}
  \Tm = \Tc - a^{-1}\left[3b\psiz^{2} + \kappa\qm^{2} + MK_{1}(h\qm)\right].
\end{equation}
Since the product $h\qm$ is a function of $\gamma$ alone, with $\qm^{2} \sim Y$ and $M \sim Y$, we see that $\Tm$ always decreases linearly with $Y$, independent of the kernel $K_{h}$.
The above two equations generalize the scaling behavior of $\Tm$ and $\Lambda$ presented in Eqs.~\eqref{main:eq:tm} and \eqref{main:eq:domain_size} of the Letter for a Gaussian kernel.

\section{Numerical techniques}
\label{sec:numerical}

As the field $\psi(\bm{x}) $ is conserved (i.e., its spatial average is a constant), we can numerically minimize the free energy
in Eq.~\eqref{eq:free_real} by considering a ``time''-dependent field
$\psi(\bm{x}, t)$ and evolving it using Model B dynamics~\cite{bray1994,provatas2011}:
\begin{equation}
  \frac{\partial \psi(\bm{x}, t)}{\partial t} = \nabla^{2}\left(\frac{\delta \mathscr{F}}{\delta \psi}\right) = \nabla^{2}\left[a(T-\Tc)\psi(\bm{x}, t) + b\psi^{3}(\bm{x}, t)
  - \kappa\nabla^{2}\psi(\bm{x}, t) + M\bar{\psi}(\bm{x},t)\right].
  \label{eq:cahn}
\end{equation}
Here $\nabla^{2}$ is the Laplacian, $t$ is the time, and $\bar{\psi}$ is the coarse-grained field defined in Eq.~\eqref{eq:cgpsi}.
We use Eq.~\eqref{eq:cahn} solely for the purposes of energy minimization.
It will not describe the actual dynamical evolution of $\psi$ seen in experiments, as it completely ignores material transport
via hydrodynamic flow and related dissipative effects~\cite{doi2009}.

Time evolution is performed after Fourier transforming Eq.~\eqref{eq:cahn} in space, which lets us
compute spatial derivatives and convolutions efficiently using fast-Fourier techniques (assuming periodic boundary conditions).
However, it is known that explicit first-order time stepping can lead to instabilities unless the time step $\delta t$ is very small.
Therefore, we use a semi-implicit method employing a linearly stabilized splitting scheme~\cite{eyre1998,vollmayr-lee2003,yoon2020}
to evolve the time-discretized version of Eq.~\eqref{eq:cahn} in Fourier space, given by
\begin{equation}
  \psi_{\bm{q}}(t+\delta t) = {C}^{-1}_{\bm{q}}\left\{{A}_{\bm{q}}\psi_{\bm{q}}(t) - {B}_{\bm{q}}[\psi^{3}(t)]_{\bm{q}}\right\}.
  \label{eq:numerics}
\end{equation}
Here $\psi_{\bm{q}}(t)$ and $[\psi^{3}(t)]_{\bm{q}}$ are the Fourier transforms of $\psi(\bm{x}, t)$ and $\psi^{3}(\bm{x},t)$,
respectively, and the coefficients ${A}_{\bm{q}}$, ${B}_{\bm{q}}$, and ${C}_{\bm{q}}$ are
\begin{equation}
  {A}_{\bm{q}} = 1 - 3a(T-\Tc)q^{2}\,\delta t,
  \quad
  {B}_{\bm{q}} = bq^{2}\,\delta t,
  \quad
  \text{and}
  \quad
  {C}_{\bm{q}} = 1 + \left[\kappa q^{2} - 2a(T-\Tc) + M\mathrm{e}^{-h^{2}q^{2}}\right]q^{2}\,\delta t.
  \label{eq:coeff}
\end{equation}
Although this method is only $\mathcal{O}(\delta t)$ accurate in time, it is sufficient for our purposes of energy minimization.
For the three-dimensional equilibrium configurations presented in Figs.~\ref{main:fig:phase}(b) and~\ref{main:fig:phase}(c) of the Letter, Eq.~\eqref{eq:numerics}
was evolved in a cubical domain of side length $10\,\text{\textmu m}$ and $256^{3}$ grid points for $5\times10^{3}$ units of time with a time step of $\delta t = 1$.
The initial field configurations $\psi(\bm{x}, 0)$ were chosen randomly, and their spatial averages were set equal to $\psiz = \phiz - \phic$.
All our numerical codes are publicly available~\cite{github}.

\end{document}